\documentclass[runningheads]{llncs}

\usepackage{graphicx}
\usepackage{tabularx} 
\usepackage{array}  
\usepackage{caption}  
\usepackage[hidelinks]{hyperref}

\usepackage{booktabs}

\usepackage{comment}
\usepackage{listings}
\usepackage{color}
 \usepackage{float} 
 \usepackage{xargs}                    
\usepackage[pdftex,dvipsnames]{xcolor}  
\usepackage{microtype}
\usepackage[colorinlistoftodos,prependcaption,textsize=tiny]{todonotes}

\usepackage[ruled,vlined,linesnumbered ]{algorithm2e}
\usepackage{algpseudocode}\usepackage{mdframed}
\usepackage{xcolor}

\definecolor{lightgray}{gray}{0.95}
\newcommandx{\change}[2][1=]{\todo[linecolor=blue,backgroundcolor=blue!25,bordercolor=blue,#1]{#2}}

\newmdenv[
  backgroundcolor=lightgray,
  linecolor=gray,
  linewidth=0.5pt,
  roundcorner=30pt,
  innertopmargin=5pt,
  innerbottommargin=5pt,
  font=\scriptsize,
]{highlightbox}

\begin{document}

\title{Communication Methods in Multi-Agent Reinforcement Learning}

\titlerunning{Communication Methods in Multi-Agent Reinforcement Learning}
\author{
Christoph Wittner\orcidID{0009-0004-1494-1730} }

\institute{Telecooperation Institute, Johannes Kepler University Linz, Austria\\
\email{k12045895@students.jku.at}\\}

\maketitle            

\begin{abstract}
Multi-agent reinforcement learning is a promising research area that extends established reinforcement learning approaches to problems formulated as multi-agent systems. Recently, a multitude of communication methods have been introduced to this field to address problems such as partially observable environments, non-stationarity, and exponentially growing action spaces. Communication further enables efficient cooperation among all agents interacting in an environment. This work aims at providing an overview of communication techniques in multi-agent reinforcement learning.  By an in-depth analysis of 29 publications on this topic, the strengths and weaknesses of explicit, implicit, attention-based, graph-based, and hierarchical/role-based communication are evaluated. The results of this comparison show that there is no general, optimal communication framework for every problem. On the contrary, the choice of communication depends heavily on the problem at hand. The comparison also highlights the importance of communication methods with low computational overhead to enable scalability to environments where many agents interact. Finally, the paper discusses current research gaps, emphasizing the need for standardized benchmarking of system-level metrics and improved robustness under realistic communication conditions to enhance the real-world applicability of these approaches.

\keywords{Machine learning \and MARL \and Communication}
\end{abstract}

\section{Introduction}
\subsection{Relevance of Communication in MARL}

% Short intro to RL
Reinforcement learning (RL) is a concept in machine learning that focuses on enabling an agent to learn the solution to a task at hand by repeatedly interacting with its environment and receiving a reward signal based on the actions it takes. The agent tries to arrive at an optimal solution policy by maximizing the discounted cumulative reward received when acting based on observations of the environment \cite{marl-book}. 

% extension to MARL
However, the concepts introduced in RL are limited to a single agent interacting with the environment, which severely restricts the space of problems to which this strategy is applicable. This drawback has led to the introduction of multi-agent reinforcement learning (MARL), which extends RL to multiple agents that simultaneously interact with the environment and learn to solve a task, either by cooperating with each other, via competitive behavior, or by a combination of the two \cite{marl-book}.  

% need for communication in MARL
Introducing additional agents to the environment gives rise to a set of new problems such as non-stationarity, exponential growth of the joint action space, and aligning individual agents' actions among each other \cite{marl-book}. These challenges have led to the formalization of different training and execution paradigms: fully decentralized (both training and execution are local) \cite{amato2024dte}, fully centralized (training and execution for all agents take place on one central node) \cite{amato2024ctde}, and centralized training with decentralized execution (CTDE) \cite{amato2024ctde}. CTDE is the most commonly applied approach, as it aims to solve the non-stationarity problem by allowing access to global information during training, while maintaining the required scalability and robustness of decentralized execution. However, in the absence of a global critic or centralized coordination, effective coordination under the CTDE framework still relies heavily on efficient information sharing. Another key issue in the MARL framework is partial observability, which describes the problem that every agent might only be able to obtain information about a small area of the environment. This hinders the agents from learning optimal policies \cite{guestrin2002coordinated}. To address these problems, agents in MARL environments are equipped with mechanisms to communicate with each other, enabling agents to efficiently cooperate and coordinate their actions, resulting in stronger performance.

% need for comm comparison 
The literature contains a large number of communication methods, ranging from implicit communication, where coordination is inferred from cues placed in the environment, to explicit frameworks such as fully connected message passing, attention-based, graph-based, and role-based communication. None of these approaches serves as a one-fits-all solution. Which method leads to the best results heavily depends on the specific characteristics of the problem at hand. For instance, while fully connected message passing ensures that information flows between all agents, it may introduce redundancy and noise in large-scale systems compared to attention-based or graph-based methods that selectively filter interactions and are therefore more computationally efficient \cite{niu2021multi}. Similarly, role-based communication can enforce the necessary structure for specialized tasks but may lack the flexibility of implicit mechanisms in dynamic, homogeneous environments. Therefore, a comparative analysis of these communication methods is essential not only to compare task performance, but to systematically analyze their inherent strengths and weaknesses to specific problem types. Such an evaluation is important for identifying which coordination strategies are best suited to the unique demands of a task, moving the field toward a better understanding of how different communication mechanisms fit distinct multi-agent challenges. Addressing this gap, this work presents a comprehensive literature study, reviewing the evolution and state-of-the-art of communication in MARL. The primary objective is to answer the question: "What cooperation mechanisms are deployed in multi-agent reinforcement learning, and what are their strengths and weaknesses?", thereby establishing a systematic comparison of current methodologies.

\section{Methodology}
The goal of this research is to provide an overview of the development and effectiveness of coordination strategies in Multi-Agent Reinforcement Learning. To support a thorough and balanced analysis, a systematic literature review was carried out with a focus on identifying different cooperation mechanisms and analyzing their strengths and weaknesses. This section describes the methodological framework used to select, categorize, and qualitatively assess the relevant literature.

\subsection{Foundation and Search Strategy}
The research process began with an exploration of the initial work that shaped the field. Investigation of early papers centered on the important contributions of Ming Tan \cite{tan1993multi}, particularly his discussion of the distinction between independent and cooperative agents, as well as influential studies by N. R. Jennings \cite{jennings1996coordination} and the overview of communication frameworks shown by L. Panait and S. Luke \cite{panait2005cooperative}. Together, these works established the historical context for understanding how communication emerged as a central requirement for addressing decentralized, partially observable Markov decision processes (Dec-POMDPs).

Building on this theoretical foundation, the literature search was expanded to capture the current state of the art. Relevant publications from leading academic journals and conferences, including NeurIPS, AAMAS, and IEEE, published between 2016 and 2025 were collected. In addition, works from arXiv were included to capture emerging trends, as a lot of novel machine learning research is accessible there before being formally published. Through this iterative search of relevant publications, a final set of 29 peer-reviewed papers was identified for detailed examination, on which all of the following sections are based.

\subsection{Research Question}
The primary research question this survey is based on was derived from the gaps observed during the initial review phase. \newline

RQ: What cooperation mechanisms are deployed in multi-agent reinforcement learning, and what are their specific strengths and weaknesses?
 \newline
 
Rather than providing a simple list of existing methods, this question is intended to enable a direct comparison of communication methods applied in MARL, giving the reader a solid understanding of the different paradigms and their applicability.

\subsection{Annotation and Classification Framework}
To offer a structured comparison of the selected literature, a systematic annotation and classification process was applied to all 29 papers. Each work was categorized based on both its primary type of contribution and its underlying communication paradigms.

With respect to the type of contribution, the works were classified as surveys, algorithms, frameworks, or applications. Survey papers were classified as "no-comm" since they show a multitude of different approaches rather than focusing on a specific type of communication. For publications not classified as surveys, the communication type was additionally determined. This labeling includes message-based communication, implicit communication, attention-driven or selective communication, graph-based relational message passing, role- or hierarchy-based coordination, and niche approaches. Many, especially more modern approaches, do not follow one of these frameworks directly but rather combine multiple of these categories. To capture these hybrid mechanisms, each work could receive multiple communication labels.

\begin{figure}
    \centering
    \includegraphics[width=1\linewidth]{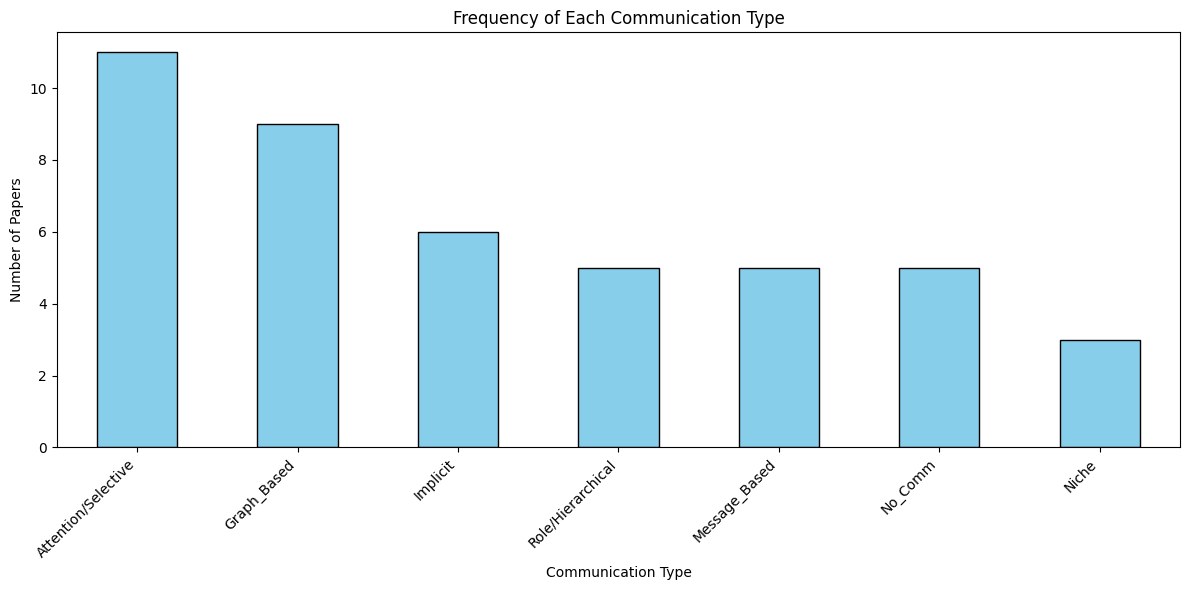}
    \caption{Communication mechanism frequency}
    \label{fig:comm_freq}
\end{figure}

\begin{figure}
    \centering
    \includegraphics[width=0.5\linewidth]{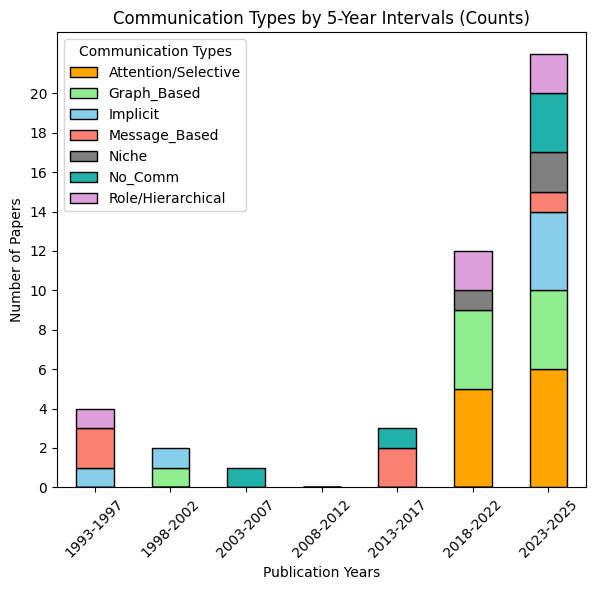}
    \caption{Communication mechanism publications over time}
    \label{fig:comm_over_time}
\end{figure}

After the labeling process, simple paper statistics were obtained to get a quick overview of the selected literature. Figure \ref{fig:comm_freq} shows that attention- and graph-based methods were included slightly more frequently than other methods and, as expected, niche approaches appear less regularly. In figure \ref{fig:comm_over_time}, we can observe that recent research also focuses on attention- and graph-based methods, as well as a newly regained interest in implicit communication.

\section{Communication frameworks}
The main challenge in Multi-Agent Reinforcement Learning is changing from independent agents to collective coordination. Communication is the mechanism of choice to overcome this challenge, allowing agents to tackle partial observability by aggregating individual observations into a more complete understanding of the global state. Research in the field started with simple data exchange and has moved toward more specialized architectures designed to handle the trade-offs between obtained global information and computational complexity.
This section provides a systematic evaluation of the most important communication paradigms identified in the literature. The following analysis categorizes these frameworks based on their structural approach to information sharing, from foundational fully-connected topologies to advanced selective and hierarchical models. For each approach, both the fundamental communication mechanics and their impact on task performance and scalability are evaluated. By highlighting the specific strengths and weaknesses of the different methodologies, this section aims to clarify which coordination strategies are best suited for varying classes of multi-agent problems.

\subsection{Fully-connected message passing}
Fully-connected message passing represents one of the foundational paradigms for communication in MARL. In this framework, agents share information via a fully connected topology, ensuring that messages are transmitted between all agent pairs. This unrestricted flow allows every agent to aggregate local observations into a wider view of the environment, effectively mitigating partial observability and enabling cooperative task solving. 

Initial research in multi-agent systems (MAS) relied on predefined message structures, often derived from classical symbolic logic, to facilitate coordination \cite{dorri2018multi}. However, these static protocols are insufficient for the dynamic, high-dimensional state spaces of reinforcement learning. A significant advancement beyond predefined message structures was introduced by Sukhbaatar et al. with CommNet \cite{sukhbaatar2016learning}, in which agents learn to generate continuous communication signals through backpropagation. By leveraging the differentiable nature of neural architectures, these frameworks allow the communication channel itself to be optimized end-to-end, resulting in significant improvements over static messages. Forster et al. \cite{foerster2016learning} refined this concept further by utilizing techniques that use continuous messages during centralized training to enable gradient flow, while discretizing those signals during decentralized execution to satisfy real-world bandwidth constraints.

However, fully connected message passing exhibits significant scalability challenges, as the cumulative communication overhead grows rapidly with the number of agents, creating computational bottlenecks and hindering the feasibility of decentralized execution \cite{niu2021multi}. Despite these limitations, fully connected approaches consistently outperform non-communicating baselines and serve as the structural basis for the development of more advanced, selective communication protocols, further explained in the following sections.

\subsection{Implicit communication}
In contrast to explicit message passing, implicit communication enables coordination without the use of dedicated communication channels. This framework, which emerged in the MARL literature around the same time as early message-passing approaches, is often compared to mechanisms observed in social insects, such as ants, which coordinate complex group behaviors by depositing pheromones in their environment \cite{panait2005cooperative}. In a MARL context, this involves agents embedding signals within their actions or by modifying the environmental state. Other agents then learn to pick up on these cues and adapt their behavior accordingly. Because this paradigm does not require a separate communication channel, it avoids the computational overhead and bandwidth bottlenecks commonly associated with fully connected architectures. These characteristics make it computationally efficient, scalable, and directly applicable to fully decentralized problems \cite{li2023explicit}.

However, reducing communication to signals embedded in actions and environmental changes introduces distinct challenges. Unlike explicit messages, implicit signals are often ambiguous, as agents must learn to distinguish between signals intentionally placed by other agents and states that arise from standard interaction with the environment. While these methods scale well, they often achieve lower performance than explicit communication approaches, since the dual-purpose nature of actions, serving both task optimization and information transmission, can lead to sub-optimal policies \cite{li2023explicit}. Even with these downsides, implicit communication remains an important area of research, particularly in environments where explicit communication is not applicable due to physical or technical limitations.

\subsection{Attention-based / Selective communication}
To address the limitations of communicating repeated or noisy information, recent research has advanced to selective communication mechanisms. This framework is built on the intuition that not all messages are equally relevant to every agent. Instead, the importance of a message is highly dependent on the specific context and the local state of the receiving agent.

By extending the fully connected message-passing framework with an attention mechanism, similar to the mechanism used in transformer architectures, agents can learn to dynamically weight incoming messages and filter out irrelevant or noisy data. An important publication in this area is Targeted Multi-Agent Communication (TarMAC) \cite{das2019tarmac}, which introduced a sender-receiver signature system. In TarMAC, agents send a \textbf{key} and a \textbf{message}, and receiving agents use a \textbf{query} to determine the relevance of that message to their current task. This attention mechanism ensures that agents focus their resources on the most important information available, which consistently leads to improved coordination and higher performance compared to standard message passing methods.

While TarMAC focuses on weighting received messages, Attentional Communication (ATOC) \cite{jiang2018learning} introduces a hard attention mechanism to decide whether communication is necessary at all. In this approach, agents use an attention unit, implemented as a self-attention layer over their own local observations, to evaluate whether communication is necessary in their current state. If this is the case, the agent initiates the formation of a local communication group with agents in a certain proximity. This dynamic creation of groups of agents transforms a dense, fully connected topology into a sparse communication graph, thereby reducing the amount of incoming messages.

However, switching from simply accumulating all incoming messages to this more selective approach comes at a high cost. The addition of attention modules, which requires calculating pairwise relevance scores (queries and keys) for all agent interactions, results in substantial computational overhead during both training and execution. While this framework effectively handles the problem of redundancy and noise seen in fully connected topologies, this increase in computation further hinders scalability to environments with a large number of agents. Despite these trade-offs, selective communication is a state-of-the-art method for solving complex environments where agents have to prioritize the most important available information.

\subsection{Graph based communication}
The process of dynamically organizing agents into local groups, as applied in ATOC, points towards a structural solution to the scalability challenges observed in earlier approaches by modeling the multi-agent system as a communication graph, based on which agents share information with each other. In this framework, agents are represented as nodes, while edges define the allowed communication channels between them. The topology of this communication graph can be \textbf{predefined} based on prior knowledge of the task or fixed environmental structures \cite{shen2021graphcomm}, \textbf{proximity-based}, where agents only interact with neighbors within a certain radius \cite{wang2023ac2c}, or \textbf{dynamically learned} to obtain task-specific communication connections optimized during training \cite{niu2021multi}.

To process information within the communication graphs, this framework often applies Graph Convolution Networks (GCNs) \cite{xu2021learning} or Graph Attention Networks (GATs) \cite{niu2021multi}. These architectures enable agents to aggregate information from their neighbors via relational message passing, making direct use of the provided graph structure. Graph-based approaches are often combined with the previously introduced attention mechanisms to again enable agents to focus on the most relevant information. However, restricting communication to a subset of agents can reduce the amount of information each agent has access to regarding the full state of the environment. Wang et al. introduce a two-hop communication strategy to extend the effective communication range, enabling agents to obtain information from other agents, to which they are not directly connected, in the communication graph \cite{wang2023ac2c}. This is achieved by using intermediate agents as relay nodes to pass messages between more distant agents. Graph-based approaches directly address the computational overhead of previous methods by constraining communication to sparse subsets of agents. By limiting interactions to relevant neighbors, communication can be scaled to environments containing a larger number of agents, thereby enabling large-scale cooperation.

Nevertheless, using a sparse graph structure introduces a direct trade-off between the scope of information available to each agent and computational overhead. While fully connected topologies provide an immediate, global view of the system, communication restricted to a local subset of agents, even when extended with multi-hop mechanisms, can delay access to important information only available further away in the communication graph. But even with these drawbacks, graph-based communication is a significant advancement in MARL, offering a solid framework for balancing the benefits of information sharing with the physical and computational constraints of large-scale multi-agent systems.

\subsection{Role based / hierarchical communication}
Role-based and hierarchical communication frameworks address scalability and coordination challenges in MARL by decomposing complex global objectives into smaller and easier solvable subtasks. This decomposition is typically achieved through a hierarchical decision-making structure or by assigning specialized roles to individual agents, enabling them to focus on specific aspects of the overall task \cite{zang2023automatic} \cite{zheng2024multi}. By promoting specialization, agents can concentrate their policy learning and communication on their assigned part of the problem, reducing learning complexity. Agents assigned to identical or closely related roles can accelerate their training by sharing learning experiences or parameters, leading to more efficient convergence \cite{wang2020roma} \cite{shao2022self}. Moreover, structuring problems around reusable subtasks enhances transferability, since agents trained to solve particular subtasks can be applied to new environments where these components are still relevant. In the new environment, the system only needs to learn how to reuse or adapt existing policies to achieve the global objective \cite{shao2022self}.

The roles and hierarchical layers used in these approaches can again be \textbf{predefined} using prior knowledge about the task or domain \cite{wang2020roma}, or they can be \textbf{learned} dynamically during training to allow task-specific structural optimization \cite{zang2023automatic}. Communication within these systems is typically constrained to reflect the imposed structure, either by limiting interactions to specific groups of agents sharing a common role or objective \cite{shao2022self}, or by handling communication implicitly through the hierarchy itself, where higher-level agents aggregate information from lower-level agents for their decision making, and lower-level agents adapt their behavior to actions taken by higher-level agents \cite{zheng2024multi}. By explicitly organizing agents into functional groups, these approaches can be applied to scenarios with dynamic numbers of agents and evolving communication topologies, where the set of agents and permissible communication links may change over time \cite{shao2022self}. This structured control of information flow reduces unnecessary message exchange while maintaining coordination, making role-based and hierarchical communication a powerful framework for scaling MARL to complex, dynamic multi-agent environments.

\subsection{Niche approaches}
\subsubsection{Communication with delay} 

Imperfections in the communication of agents are an important but so far less studied aspect of communication in MARL. Ikeda et al. target agent communication where messages are not received immediately, but only after a fixed delay \cite{ikeda2022centralized}. To tackle this challenge, their approach makes use of gated recurrent units (GRUs) to accumulate information from multiple delayed messages over time, enabling agents to base decisions on temporally aggregated information.  

\subsubsection{Language-grounded communication}

The communication protocols learned by agents in MARL are often difficult to interpret, especially in better-performing frameworks that rely on emergent rather than predefined messages. Li et al. address this limitation by introducing language-grounded communication. In this approach, agents receive an additional supervised communication signal during training, which is used to guide the learned messages toward alignment with natural language. This alignment enables humans to easily interpret how agents make use of the communication channel, and further allows for zero-shot interaction with newly added agents or even with humans interacting with the environment. 

\subsubsection{Tacit communication}

Li et al. introduce tacit communication to MARL as a more recent extension of implicit communication approaches \cite{li2023explicit}. During training, agents initially learn to cooperate through explicit attention-based communication, achieving strong performance on the given task. Over time, the amount of direct communication available to agents is gradually reduced, encouraging them to encode the information they need to share within the environment itself. After completion of the full training process, direct communication has completely disappeared, and coordination is based only on implicit signals. Importantly, performance remains comparable to that achieved with message-based communication while enabling fully decentralized execution, as long as the agents' trajectories are not entirely independent.

\section{Discussion and Conclusion}

\subsection{Future Work}
A lot of progress has been made in developing communication mechanisms for MARL, but multiple important issues remain open for future investigation. A key gap observed in this survey is the lack of a comprehensive and standardized comparison of communication approaches with respect to practical system-level metrics. Existing work mainly evaluates methods based on task performance, such as reward or convergence speed, while metrics like communication overhead, runtime complexity, bandwidth usage, memory requirements, and scalability with increasing agent counts are rarely analyzed in a standardized way. Future research would benefit from systematic benchmarking frameworks that evaluate these metrics for explicit, implicit, attention-based, graph-based, and hierarchical/role-based communication paradigms under standardized experimental conditions. This would enable practitioners to more easily compare the trade-offs of different approaches and provide guidance for selecting optimal communication strategies in real-world applications.

Another important research direction is the robustness of communication under non-ideal conditions. Most MARL communication frameworks assume reliable, instantaneous, and noise-free message exchange, an assumption that rarely holds in real-world multi-agent systems. Expanding research on faulty or imperfect communication, such as delayed, lossy, noisy, or partially corrupted messages, remains essential to improve real-world applicability. Early work has started to explore delayed communication, but more studies are needed to understand how different communication paradigms are impacted by realistic constraints and how agents can learn to adapt their coordination strategies accordingly, thus improving real-world applicability.

In general, future MARL communication research will require shifting focus beyond algorithmic performance toward a more holistic evaluation of robustness, efficiency, and deployability. Addressing these open challenges would improve the practical relevance of multi-agent reinforcement learning and support its application to complex, real-world domains.

\subsection{Conclusion}

Communication in MARL is a highly relevant and rapidly evolving area of research, enabling agents to coordinate effectively in increasingly complex multi-agent environments. This work compares explicit, implicit, attention-based, graph-based, and hierarchical/role-based communication approaches, highlighting their respective strengths, weaknesses, and trade-offs. In particular, the survey shows the central role of communication efficiency in achieving scalability as the number of agents and the task complexity increase. Since there is no single best-performing solution, the choice of communication mechanism strongly depends on the environmental structure, computational constraints, and execution requirements. Finally, this work outlines several promising directions for future research, especially aimed at improving robustness and practical applicability under realistic communication constraints.

\subsection{Acknowledgments}
\textbf{Use of AI}
The conceptual and factual content of this paper was created without the use of any AI-tools. AI was only used for spell-checking and linguistic refinement.

\bibliographystyle{splncs04}
\bibliography{references}
\end{document}